**Optical Metamaterials at Near and Mid IR Range Fabricated by Nanoimprint Lithography**

Wei Wu[1*†], Evgenia Kim[2*], Ekaterina Ponizovskaya[1], Yongmin Liu[3], Zhaoning Yu[1], Nicholas Fang[4], Y. Ron Shen[2], Alexander M. Bratkovsky[1], William Tong[1], Cheng Sun[3], Xiang Zhang[3], Shih-Yuan Wang[1] and R. Stanley Williams[1]

[1] Quantum Science Research, Hewlett-Packard Laboratories, Palo Alto, CA 94304

[2] Department of Physics, University of California, Berkeley, CA 94720

[3] NSF Nano-scale Science and Engineering Center (NSEC), University of California, Berkeley, CA 94720

[4] Department of Mechanical & Industrial Engineering, University of Illinois, Urbana-Champagne, IL 61801

**Abstract.**

Two types of optical metamaterials operating at near-IR and mid-IR frequencies, respectively, have been designed, fabricated by nanoimprint lithography (NIL), and characterized by laser spectroscopic ellipsometry. The structure for the near-IR range was a metal/dielectric/metal stack "fishnet" structure that demonstrated negative permittivity and permeability in the same frequency region and hence exhibited a negative refractive index at a wavelength near 1.7 μm. In the mid-IR range, the metamaterial was an ordered array of four-fold symmetric L-shaped resonators (LSRs) that showed both a dipole plasmon resonance resulting in negative permittivity and a magnetic resonance with negative permeability near wavelengths of 3.7 μm and 5.25 μm, respectively. The optical properties of both metamaterials are in agreement with theoretical predictions. This work demonstrates the feasibility of designing various optical negative-index metamaterials and fabricating them using the nanoimprint lithography as a low-cost, high-throughput fabrication approach.

[*] The first two authors contributed equally to this work.

[†] email: wei.wu@hp.com

**Introduction**

Lord Raleigh's (1877) work on waves in dispersive media demonstrated the possibility of a negative index of refraction and backward waves [1]. In such a case, the phase velocity would be opposite to the group velocity in a certain frequency range with strong dispersion (near a particular excitation of the medium, or some resonant transition, and in certain types of photonic crystals). Pafomov[2] and Veselago[3] showed that backward waves can propagate in an isotropic medium with both negative permittivity $\varepsilon$ and permeability $\mu$ and, consequently, a negative refractive index (Negative Index Materials, or NIM). Early calculations showed that such backward waves can propagate in plasma-like bodies, implying that the phase shift of the transmitted electromagnetic wave propagating through a NIM system will be negative too. R.A. Silin (1966) has proved that backward waves can be supported by artificial periodic structures, now called metamaterials. Recent interests in NIM systems followed theoretical predictions of negative permittivity and permeability in periodic metallic metamaterials of interspersed metallic meshes producing $\varepsilon < 0$ [4] and split-ring resonators (SRR) that have magnetic resonance and provide $\mu < 0$ [5]. Interest in NIM became very intense after first speculation by J.Pendry that even metamaterials with only negative permittivity can produce an image of sources of magnetic fields with a sub-wavelength resolution[6]. This result has serious limitations [7], but it was indeed possible to demonstrate a Vesalago lens with sub-wavelength resolution with just a silver slab, as originally suggested by Pendry, and certain photonic crystals. The importance of artificial negative index metamaterials becomes apparent when considering that the natural materials do not show strong magnetic responses, especially negative magnetic permeability above terahertz frequencies. Pendry proposed fabricating split-ring resonators (SRRs) to engineer the effective permeability $\mu_{eff}$ by means of inducing eddy currents in the rings with strong diamagnetic response (the material used to fabricate SRRs for such a magnetic response is actually a nonmagnetic good conductor).[8] From the point of view of equivalent circuit theory, a SRR is an

*RLC* circuit with a standard resonance frequency given by $\omega_0 = 1/\sqrt{LC}$, where *L* is the inductance and *C* is the capacitance of the SRR structure, respectively. At frequencies near $\omega_0$, the magnetic flux penetrating through SRRs induces a circulating current, giving rise to an effective magnetic dipole that produces a field opposite to the external one. The magnetic dipole can respond in phase (*i.e.*, $\mu_{eff} > 1$) or out of phase (*i.e.*, $\mu_{eff} < 1$) with respect to the external field. A negative $\mu_{eff}$ can be achieved if the resonance strength is sufficiently strong (that is, the *Q* factor of the equivalent RLC circuit is sufficiently large). With advances in nanofabrication techniques that have enabled the patterning of sub-optical wavelength features, researchers have been able to implement artificial magnetic metamaterials from terahertz to telecommunication frequencies,[9-12] and even in the visible range[13,14].

In this paper, we present theoretical and experimental results of two types of meta-structures, the periodic hole array trilayer or "fishnet-structure"[13] and a novel magnetic L-shaped resonator (LSR). The former showed the existence of both negative magnetic permeability and negative dielectric permittivity at the same near-IR wavelength region, and hence a negative refractive index. The latter showed the presence of plasmon and magnetic resonances in the mid-IR range. Both of these structures used magnetic resonance to achieve negative magnetic permeability.

The "fishnet" structure that we used to demonstrate a negative refractive index at near-IR wavelengths is illustrated in Fig.1 (a), (b). It consists of two metallic fishnets (metallic films with a regular hole array) separated by a dielectric layer[13]. The wide metallic wires create a negative permittivity by a mechanism similar to that for periodic metallic metamaterials[4] (soft collective plasmon-like excitation). The magnetic field H penetrating areas between the layers induces a magnetic response, as schematically shown in Fig.1 (b), that results in $\mu_{eff} < 0$ (magnetic "plasmon"). In our design, we used two 25 nm thick Ag films with a 35 nm thick

SiO$_2$ spacer layer in between. The widths of the metallic wires that composed the "fishnet" were 100 nm in one axis and 300 nm in the other. This system was designed to have a negative index at λ=1.5μm with the use of full scale FDTD simulations [15].

A unit cell of the LSR meta-structure consisted of four isolated L-shaped gold arms with four-fold rotation symmetry. The geometry and dimensions of our LSR design are shown in Fig. 2. LSRs possess four-fold rotational symmetry therefore it is expected that the electric coupling cannot form a net circulating current, leading to a magnetic dipole moment. More specifically, for an external electromagnetic wave that propagates along the *z*-axis, the in-plane electric field (polarizing along *x*-axis) would drive the vertical two L-shaped arms identically (without considering retardation since the structure is in sub-wavelength scale).

Magnetic fields can be induced by the charge oscillation in the vertical arms at Mie resonance. However, the total magnetic moment within the unit cell is zero, because of the symmetric arrangement of the four L-shaped arms. In other words, the in-plane electric fields have little contributions to the effective magnetic moment. Consequently the bianisotropic response to the in-plane *k*, which is present in traditional SRRs, is suppressed, because bianisotropy arises when the incident electric fields could couple to SRRs and induce the magnetic resonance; and vice versa, magnetic fields can result in electric resonances[16,17]. The bianisotropy is undesirable because the electromagnetic constitutive parameters are highly dependent on the orientation of SRRs with respect to the excitation field. By introducing structures with high symmetry like LSRs, the bianisotropy could be suppressed or eliminated[18].

Aside from the ideal geometry, the fabrication of those meta-structures with high-precision, high-throughput, and low-cost posed another challenge, especially for metamaterials operating in the infrared or visible range: nanoscale patterning resolution is required, since these feature sizes are smaller than the resolution of state-of-the-art photolithography due to the diffraction limit. Most of the previous works at this range have been fabricated by either electron-beam lithography (EBL) [9,19,20] or focused ion beam (FIB), However, both EBL and FIB have very low throughput and consequently are not feasible for mass production. The semiconductor industry has worked on the next generation lithography (NGL) for several decades in order to extend Moore's law. As one of the NGL candidates, nanoimprint lithography (NIL)[21] performs pattern transfer by the mechanical deformation of the resist rather than photo- or ion-induced reaction in the resist as for most other lithographies. The resolution of NIL is thus not limited by the wavelength of the light source. Moreover, NIL is a parallel process; therefore it has a high throughput. NIM fabrication, which requires high patterning resolution, does not have stringent requirements for overlay or low defect levels, which are considered key challenges for NIL.[22,23] Thus NIM is an ideal application for NIL, and it was successfully used to fabricate L-shaped resonator metamaterial [24].

**Simulation**

The effective electromagnetic parameters of the "fishnet" structure were numerically calculated using the Finite Difference Time Domain (FDTD) method[25] (for details of the model and calculations see ref. 15). The dielectric permittivity of Ag was described by the standard Drude model. As an intrinsic feature of metallic NIM, surface plasmons are excited at the interfaces between the metal and vacuum and the metal and dielectric[26]. The incident radiation field induces currents at the metallic surfaces that lead to losses. High losses degrade the resonance strength, hence preventing the refractive index from reaching negative values. Our structure was designed to have minimal losses. The dielectric layer thickness and the width of the wires had a strong influence on the position of the magnetic resonance, which occurred in our case in the wavelength range with $\text{Re}[\varepsilon_{eff}] < 0$. The optimal structure was obtained when the magnetic permeability $\text{Re}[\mu_{eff}] < 0$ was very close to the edge of the range where $\text{Re}[\varepsilon_{eff}] < 0$

(see Fig. 3 (a),(b)). At longer wavelengths, Re[$\varepsilon_{eff}$] is more negative and Im[$\varepsilon_{eff}$] increases; since Im[$\varepsilon_{eff}\mu_{eff}$] increases, and the loss becomes more significant. The magnetic resonance leads to a negative Re[$\mu_{eff}$] near a wavelength of 1.65 μm, while Re[$\varepsilon_{eff}$] has the typical metallic behavior of changing monotonically from positive to negative values with a crossover wavelength around 1.45 μm in our design. To calculate the refractive index (n), we have used the method that retrieves n from the transmission and reflection coefficients and their respective phase shifts[27].

For the LSR array, we also calculated $\varepsilon_{eff}$ and $\mu_{eff}$ using Microwave Studio by CST, which is a commercial electromagnetic solver based on the finite difference time domain (FDTD) algorithm and the standard Drude model:

$$\varepsilon(\omega) = 1 - \frac{\omega_p^2}{\omega(\omega + i\gamma)},$$

where $\omega_p = 1.367 \times 10^{16}$ rad/s is the bulk plasma frequency and $\gamma = 4.07 \times 10^{13}$ rad/s is the electron collision frequency of the gold in the infrared region. We first simulated the complex transmission and reflection coefficients of a single LSR unit with open boundary conditions along the wave propagation direction, and with electric and magnetic boundary conditions on the transverse direction. Then the effective electric permittivities and magnetic pemeabilities can be calculated[27].

As shown in Fig. 4 (a), there is a clear Drude-Lorentz resonance for Re[$\mu_{eff}$] at 5.22 μm. Because of the strong magnetic resonance, Re[$\mu_{eff}$] is negative over a broad region (~4.5-5.16 μm). Accompanying this magnetic resonance is an anti-Drude-Lorentz resonance in the $\varepsilon_{eff}$ at this wavelength region, even though $\varepsilon_{eff}$ never reaches a negative value. This electric anti-resonance is caused by the bounded refractive index of the structure which possesses intrinsic finite periodicity. In addition, LSRs also possess a pronounced electric resonance, which results in a negative $\varepsilon_{eff}$ at shorter wavelengths below 3.9 μm (see Fig. 4 (b)). Since the effective index $n_{eff} = \sqrt{\varepsilon_{eff}\mu_{eff}}$ has large imaginary part as $\text{Re}(\varepsilon_{eff}) \cdot \text{Re}(\mu_{eff}) < 0$, it is expected that a strong reflectance occurs at both the magnetic and electric resonance frequencies.

The advantage of the fishnet structure compared to the LSR array is that the former has a very wide range of negative $\text{Re}[\varepsilon_{eff}]$, a characteristic of the Drude response in metals, that makes it easier to combine with the narrow magnetic resonance $\text{Re}[\mu_{eff}]<0$. On the other hand, the LSR array has such narrow electric and magnetic resonances that it is difficult to have them at the same wavelengths, and therefore reach a negative value of the refractive index.

**Fabrication**

The fabrication of meta-materials by NIL involved two major steps: the NIL mold fabrication and the device fabrication. We describe the fishnet structure fabrication as an example. The NIL mold consisting of the fishnet pattern was fabricated on Si substrate with EBL, lift-off, and reactive ion etching (RIE). First, the "fishnet" pattern was generated in the resist by EBL. The positive EBL resist was 950k-molecular-weight PMMA and the developer was a mixture of MIBK and IPA (1:3). Then, 10 nm of Cr was deposited onto the resist pattern by E-beam evaporation, and a metal lift-off process was employed to dissolve away the resist and leave the Cr "fishnet" pattern on top of the Si substrate. Finally, the pattern was etched into Si by RIE using Cr as a mask. After cleaning and application of a release coating layer, the NIL mold were completed and could be used for metamaterial fabrication[22] (Fig. 5).

The "fishnet" structure was fabricated by a double-layer UV-curable NIL process.[23] There were five steps in the device fabrication (see Fig. 6). First, a transfer layer and a liquid UV-curable NIL resist layer were spin-coated onto a glass substrate. The NIL resist was imprinted with the mold using the UV-curable NIL process. The resulting pattern was transferred into the transfer layer after residue-layer and transfer-layer RIE etchings. A stack of

Ag (25 nm)/ SiO$_x$ (35 nm)/Ag (25 nm) was deposited onto the patterned resist structure using E-beam evaporation. Finally, a lift-off process was employed to remove the transfer material and the resist to leave the stack fishnet structure directly on the glass substrate. SEM images of a portion of the resulting 500 μm x 500 μm fishnet are displayed in Fig. 7 (a), (b).

The LSR arrays were fabricated by similar processes, with two major differences.[24] First, a transparent NIL mold was needed to facilitate the curing of the resist, because the LSRs were fabricated on non-transparent substrates. The silicon mold made by EBL needed to be duplicated by NIL followed by lift-off and RIE into transparent glass daughter molds, which were used subsequently in LSRs fabrication. In this approach, the transparent molds were fabricated without EBL, which requires a conductive substrate. Second, because both Si and SiO$_2$ absorb mid-IR radiation, they are unsuitable as substrates for mid-IR metamaterials, thus our LSRs were fabricated on a thin SiN$_x$ membrane (Fig. 8 (a)). SiO$_2$ and Si$_3$N$_4$ films were deposited onto the substrate by low pressure chemical vapor deposition (LPCVD) before the resist patterning. After the LSR structures were patterned by NIL (we used gold metal instead of silver), windows were opened on the back side of the wafer by a KOH anisotropic wet etch to remove the silicon, and an HF etch was used to removed the SiO$_2$ layer to leave the Si$_3$N$_4$ membrane with the metal LSR array on it. The SEM images of a portion of the fabricated 1mm x 100μm metal LSRs array are displayed in Fig. 8 (b).

**Experiment**

The optical properties of the metamaterials were characterized with an IR spectroscopic apparatus with high spectral resolution and broad spectral tuning range based on optical parametric generator and amplifier system (OPG/OPA). The tripled output of a YAG:Nd$^{+3}$ laser with fundamental wavelength 1.064 μm, 20-Hz repetition rate, and 20-ps pulse duration was used to supply the third-harmonic radiation, which is a pump for the OPG/OPA system. The OPG/OPA output could be tuned over the range of signal wavelength from 450 nm to 2500 nm.

The OPG/OPA output was used for characterizing the "fishnet" structure. After spatial filtering to improve the quality of the beam, it was focused onto the meta-structure, which had an area of 500 x 500 μm². The input and output polarizations were controlled by a calcite polarizer and analyzer, respectively. Linearly polarized light was used for reflection geometry at a small angle of incidence of $10^0 \pm 2^0$ to excite the magnetic resonance more efficiently. For the transmission geometry the beam was incident normally on the meta-structure surface as shown in Fig. 9 (a). A germanium detector was used to measure the output signal.

Broad-bandwidth infrared pulses tunable from 2.5 to 9 μm were generated by difference-frequency (DF) mixing the tunable idler pulses of the OPG/OPA with the single wavelength output of the YAG:Nd$^{+3}$ laser at 1.064 μm in a 1 cm thick AgGaS$_2$ nonlinear crystal (see Fig. 9 (c)). The size of the metamaterial area of the LSR array was about 1 mm x 0.1 mm, and the area of illuminated spot was about 80 x 250 μm² on the sample for decreasing the non-resonant background from the substrate. The IR probe beam of the linear polarized light was at an incident angle of $70^0 \pm 2^0$ which has been calculated to maximize the extent of magnetic flux to excite the magnetic resonance of LSRs. For s-polarized incident light, as the optical magnetic field H(ω) crossed over the LSRs frame (see Fig. 9 (b)), the electric and magnetic responses both could exhibit resonant behaviors depending on the excitation frequency. In the case of a p-polarized probe beam the optical magnetic field H(ω) was parallel to the plane of the LSRs, and thus only an electric response was expected. The incident polarization of the incoming beam was controlled by a set of mirrors, and the output polarization was determined by a ZnSe analyzer. The signal from the LSR array was detected by a thermoelectrically cooled photovoltaic detector.

For complete determination of the refractive index (see Eq.3)[27], phase shift between the orthogonally polarized waves due to anisotropy of the metamaterial was measured. The experimental setup is shown in the insert of Fig. 11 (a). The polarizer and analyzer were aligned in the crossed polarized position. A Soleil Babinet Compensator was used as a zeroth-order waveplate with variable phase retardation. First, the spectrum of phase shifts was measured without the sample to determine the birefringence only from the compensator. After that, the sample was set at the $45^0$ azimuthal angle (the position shown in Fig. 1(a) corresponds to $0^0$ azimuthal angle) and the phase shift spectrum of the sample was directly measured. The substrate did not contribute to the phase shift in this case. The experimental error of the phase shift measurements was $\pm 2^0$.

**Results and discussion**

Fig. 10(a) shows the results of the transmittance (black curve) and the reflectance (red curve) measurements on the "fishnet" structure. With the optical electric field polarized parallel to the thin wires and magnetic field along the thick wires (see Fig. 1(a)), a strong resonance appears in the vicinity of 1.67 µm. This resonance comes from the negative permittivity combined with the magnetic resonance. The position of the resonance agrees with predictions from simulation (see Fig. 3(a,b)). Displayed in Fig. 10(b) is the FDTD simulation that also shows overall good agreement with experimental data.

Fig. 11(a) shows results of the phase measurements for transmitted ($\delta\varphi_2$) and reflected ($\delta\varphi_1$) light. The phase shift for orthogonally polarized light reached $-55^0$ in transmittance and $10^0$ in reflectance in the resonant wavelength region of 1.50-1.74 µm. Below 1.5 µm, the phase shift decreases and reaches zero at $\lambda = 1.13$ µm. Above 1.74 µm, it decreases monotonically beyond 1.9 µm.

Fig. 11(b) shows the real part of the refractive index, $n'$, deduced from the data using Eq.3[27] (black curve) and from the FDTD simulations (red curve). Both curves display a dip at the resonance. The value of $n'$ is negative between 1.6 and 1.73 µm. In particular, at 1.7 µm, $n' = -1.55 \pm 0.25$ and $n = n' + n'' = -1.55 + i0.03$, which suggests relatively small losses.

Theoretical calculations predicted the dipole plasmon and magnetic resonances to occur at different wavelengths for the LSR array, thus, the complex value of the refractive index. Figure 12(a) shows the s- and p-polarized reflection spectra from the LSR array on the $Si_3N_4$ membrane covered with $SiO_2$. The reflection spectrum for the s-polarized light reveals two peaks. A sharp peak resulting from negative permeability $\varepsilon_{eff}$ due to plasmon resonance of LSRs was clearly observed at 3.75±0.05 µm wavelength. Another peak due to the magnetic resonance was observed at 5.25±0.05 µm wavelength. The reflectance spectrum for the p-polarized light exhibited only the plasmon resonance at 3.75±0.05 µm because the in-plane magnetic field could not excite a magnetic resonance in this case. The positions of the electric and magnetic resonances were in good agreement with predictions from the simulated effective parameters (see Fig. 4(a,b)). The s- and p-polarized reflection spectra from the $Si_3N_4$+$SiO_2$ substrate free from LSRs showed a smooth dispersive background without resonance behavior ( see insert Fig. 12 (a)).

The s- and p-polarized reflection spectra of the LSR array on the $Si_3N_4$ + $SiO_2$ supported by the Si wafer are shown in Fig. 12 (b). Both resonances were weaker and broader because of the influence of the higher effective refractive index of the $Si_3N_4$ + $SiO_2$ + Si substrate. Also, the resonance peaks display a red shift compared with the spectra from LSRs on the $Si_3N_4$ + $SiO_2$. The effective refractive index of the $Si_3N_4$ + $SiO_2$ membrane was approximatelly n~2, and that of $Si_3N_4$ + $SiO_2$ + Si was more than 2 (n~3.7 for Si). The plasmon resonance of a metal structure would be red-shifted by an increase in the dielectric constant of the material nearby. Magnetic resonances would also be red-shifted mainly because of the increasing the capacitance.

To stress the magnetic resonance and suppress the dipole plasmon resonance, we plotted the s-polarized reflection spectrum normalized against the p-polarized spectrum from the sample of LSR on a $Si_3N_4$+ $SiO_2$ substrate (see insert Fig. 12 (b)). The normalized spectrum for the $Si_3N_4$+ $SiO_2$ substrate free from LSR's appears to be nearly dispersionless, as expected.

As a further confirmation of the agreement between our theoretical models and experimental observations, reflection spectra for s- and p-polarized excitation have been calculated and compared with the data. First, the single LSR unit cell illuminated by 70° incident plane waves was simulated. Then the far field of a LSR array (80 x 250 units) was calculated with the antenna array function in Microwave Studio. After the far-field intensity to the incident light has been normalized, the reflection spectra were obtained, as shown in Fig. 12 (c). The simulated reflection spectra are in good agreement with the experimental data: the spectral positions and the widths of the resonances were very similar.

**Conclusions**

In conclusion, nanoimprint technology has been used to fabricate periodic metamaterialss for the near and mid IR ranges. For the "fishnet" structure, the negative index of $n' = -1.55 \pm 0.25$ was observed at the $\lambda = 1.7$ μm. The novel meta-material comprising an ordered array of L-shaped resonators shows both electric plasmon and magnetic resonances in the vicinity of 3.7 μm and 5.25 μm radiation, respectively. The experimental results are in good agreement with the theoretical simulations. This work demonstrates the feasibility of using the NIL technique for the parallel production of optimized optical meta-materials.

**Acknowledgements**

The authors acknowledge DARPA for partial support.

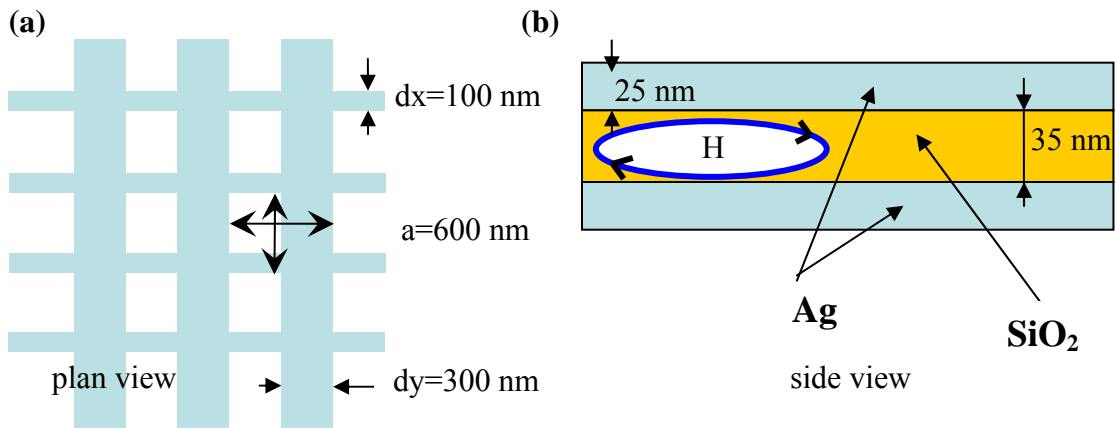

Fig. 1. A schematic of the "fishnet" metallic trilayer structure: plan view (a) and side view (b). The design parameters: 25nm Ag layers and 35 nm spacer $SiO_2$ layer.

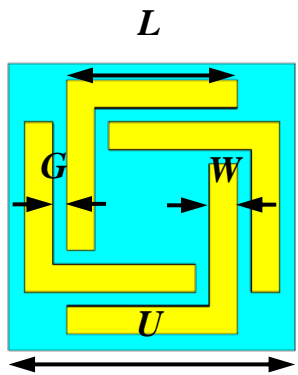

Fig. 2. A schematic of the L-shaped resonator (LSR). The design parameters: gap G = 45 nm, arm width W = 90 nm, arm length L = 550 nm, lattice constant U = 930 nm, and metal (gold) thickness t=100nm.

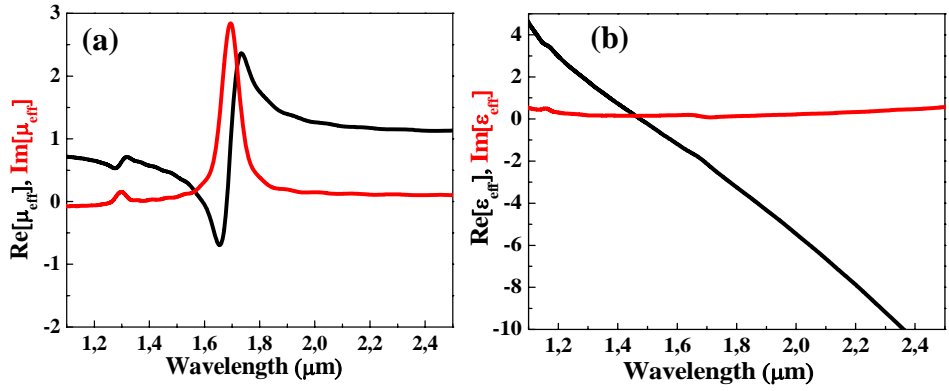

Fig. 3 (a) Simulated effective magnetic permeability ($\mu_{eff}$), and (b) effective electric permittivity ($\varepsilon_{eff}$) for the "fishnet" structure.

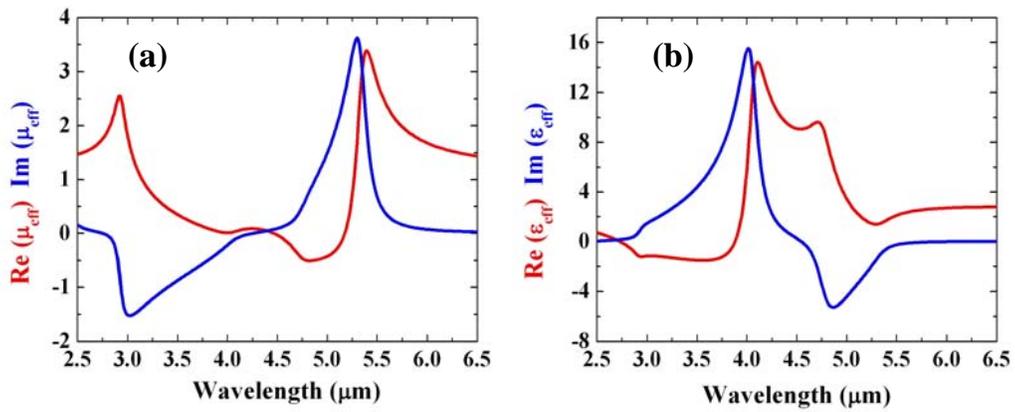

Fig. 4 (a) Simulated effective magnetic permeability ($\mu_{eff}$), and (b) effective electric permittivity ($\varepsilon_{eff}$) for the LSRs array.

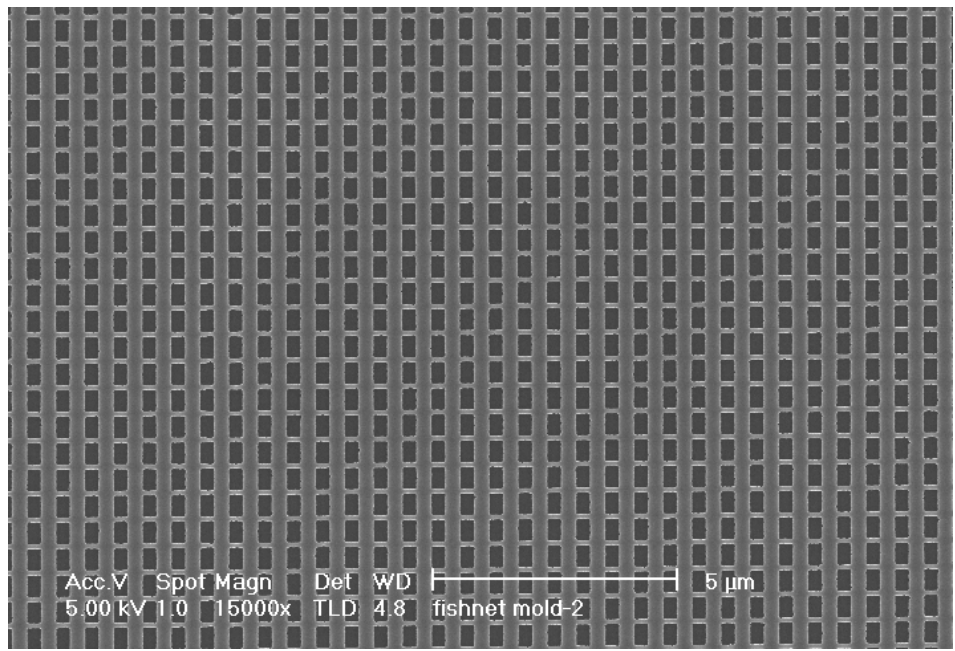

**(a)**

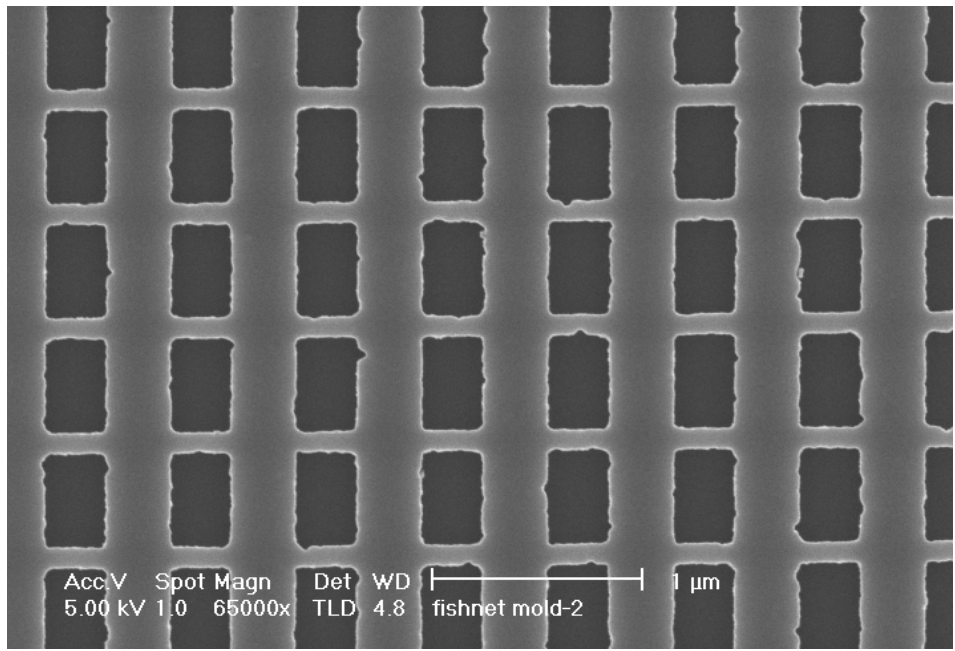

**(b)**

Fig. 5 SEM images of the NIL mold with fishnet patterns. (a) is at lower magnification than (b).

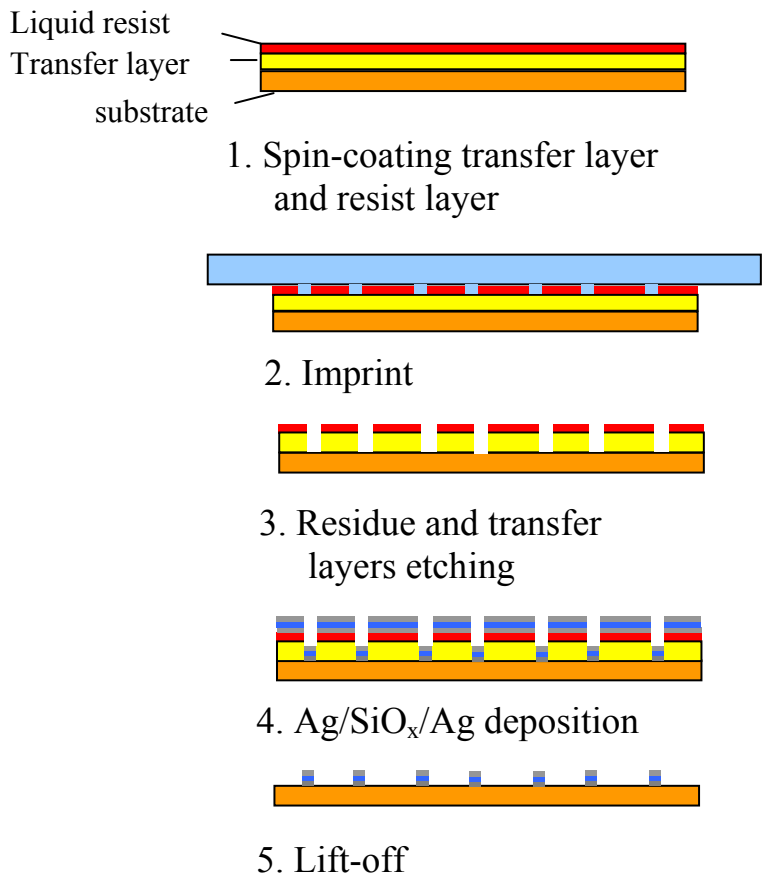

Fig. 6. Process flow of the near-IR metamaterial fabrication

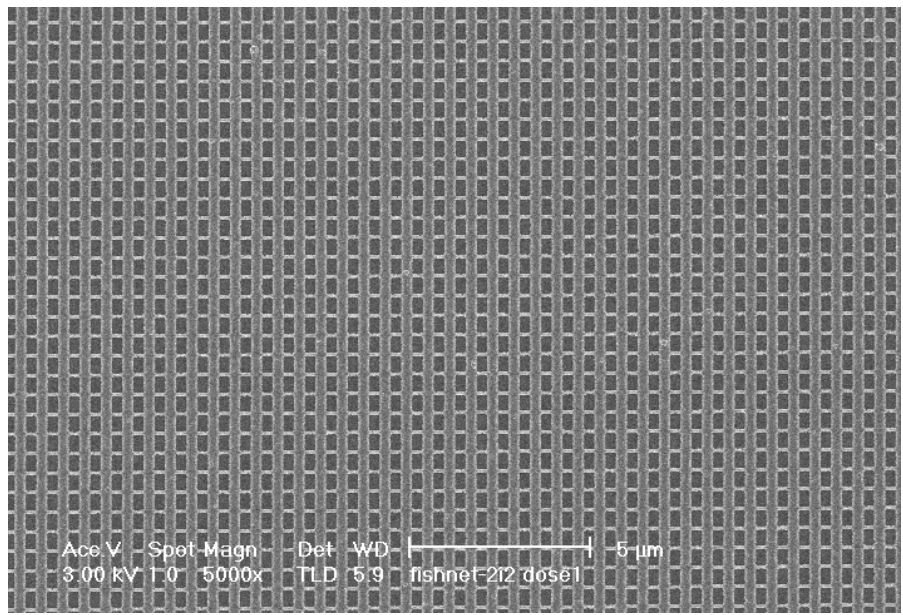

(a)

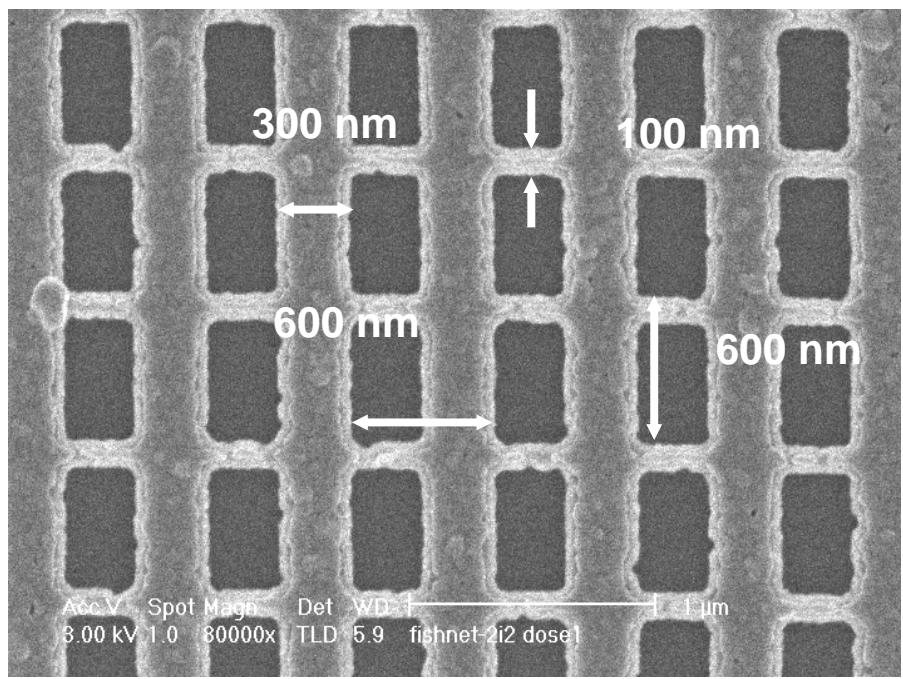

(b)

Fig. 7 SEM images of the "fishnet" structure. (a) is at lower magnification than (b).

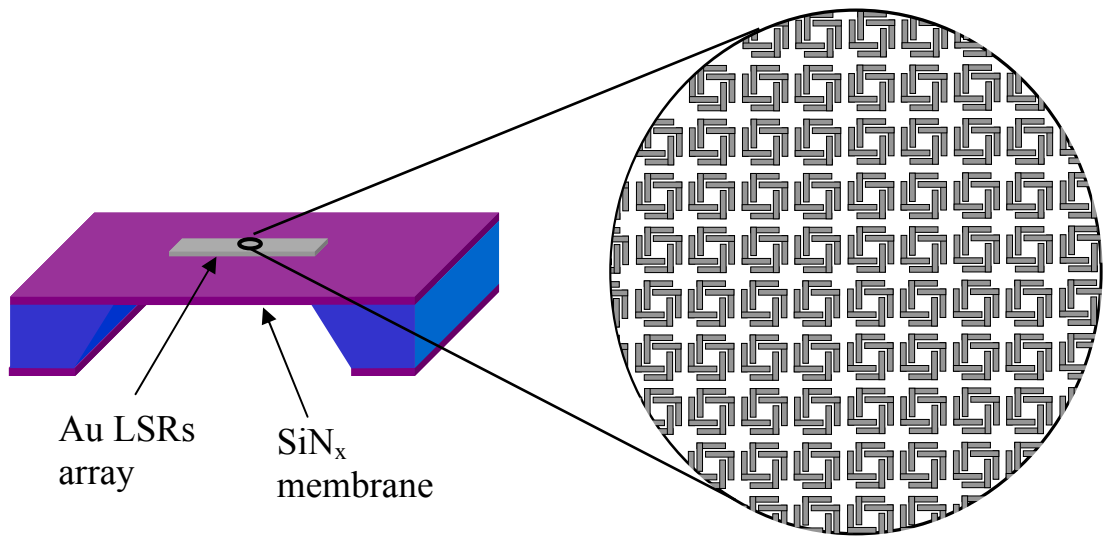

**(a)**

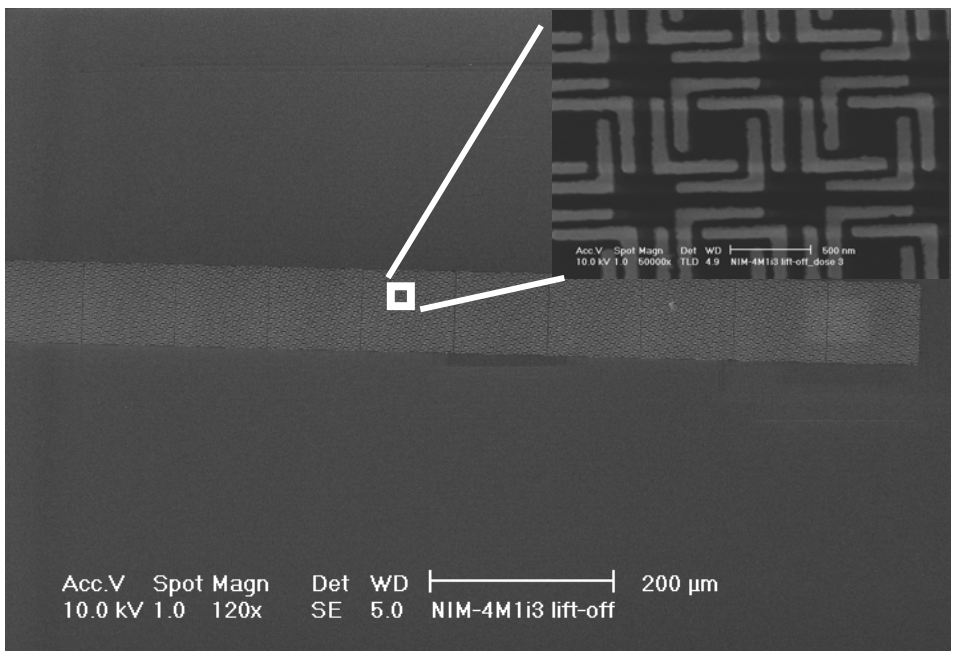

**(b)**

Fig. 8 (a) A schematic of the LSR array on $Si_3N_4$ membrane. (b) SEM images of the LSR array. The inserted image is a zoomed-in view.

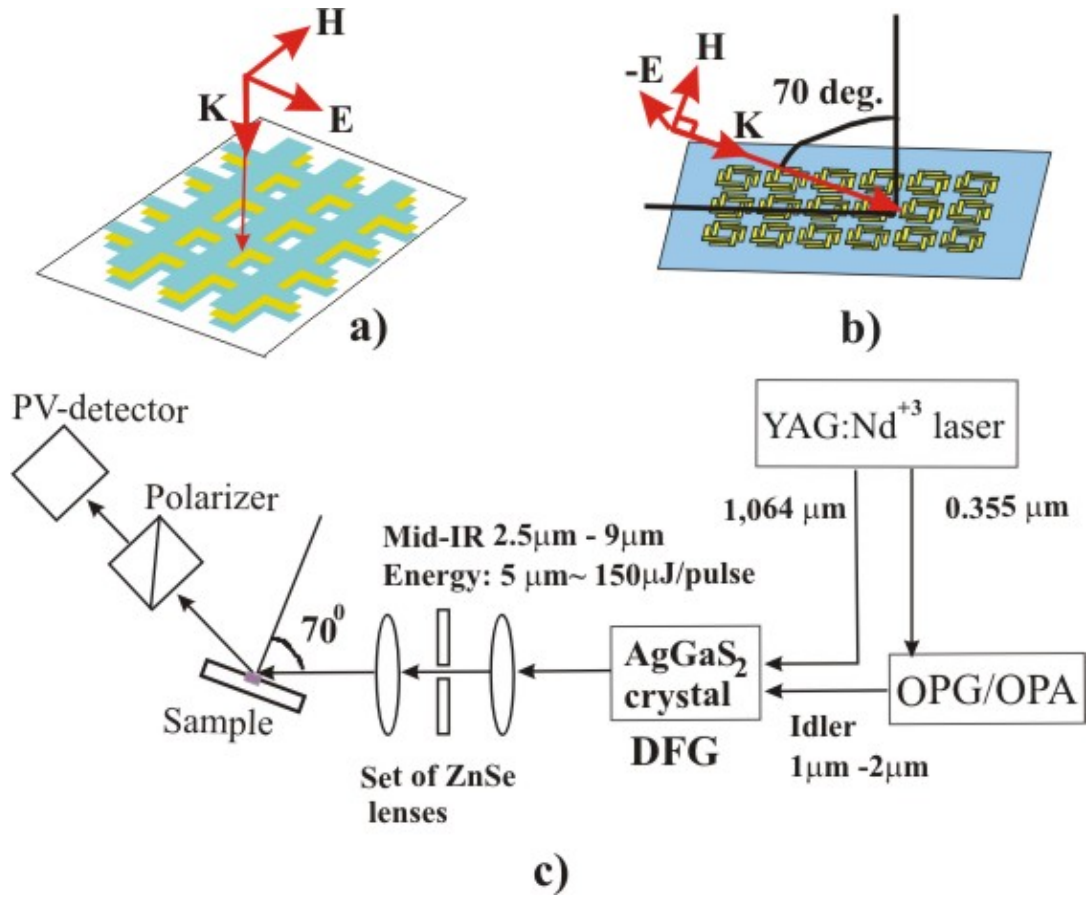

Fig. 9 (a) Illustration of the experimental configuration for the transmittance spectroscopy with probing light incident normally to the "fishnet" structure. (b) Illustration of the experimental configuration for the reflectance spectroscopy with s-polarized light incident at 70° in order to excite the magnetic response of the LSR array. (c) Schematic illustration of the experimental set up for the reflectance spectroscopy of the LSR array.

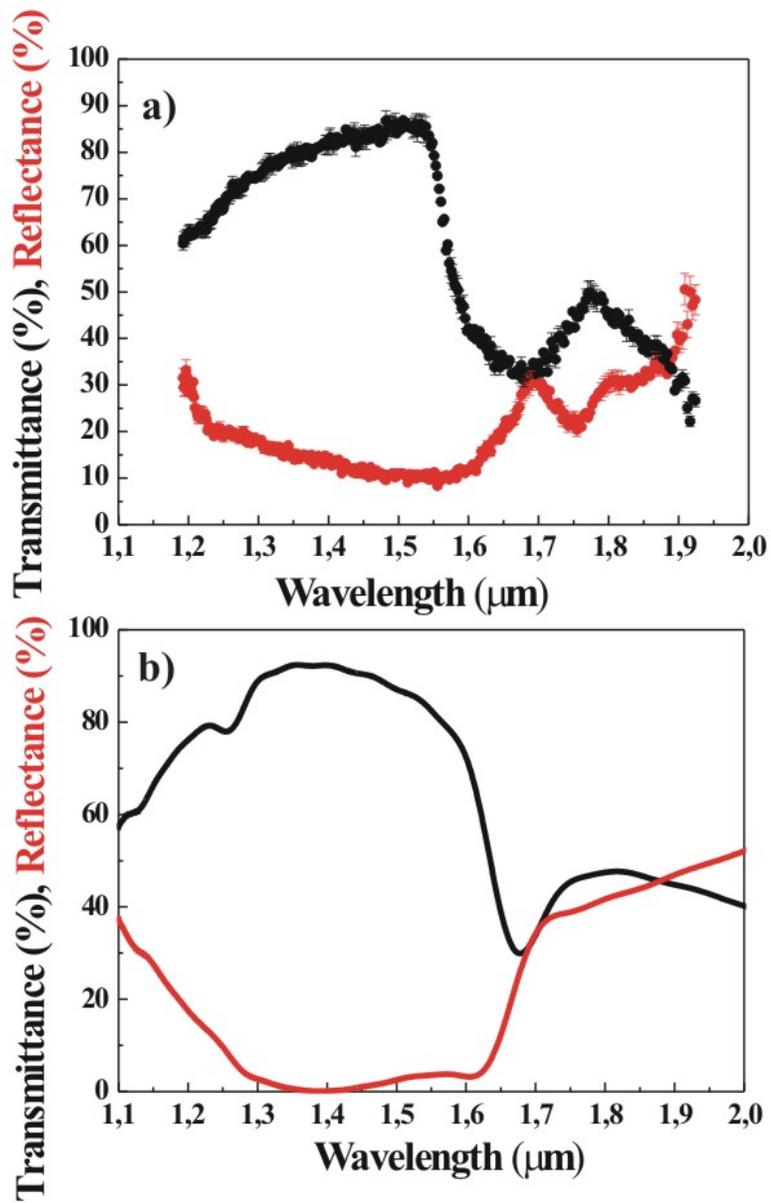

Fig. 10 (a) Transmittance (black curve) and reflectance (red curve) spectra for the "fishnet" structure. (b) Simulated transmittance (black curve) and reflectance (red curve) spectra for the "fishnet" structure.

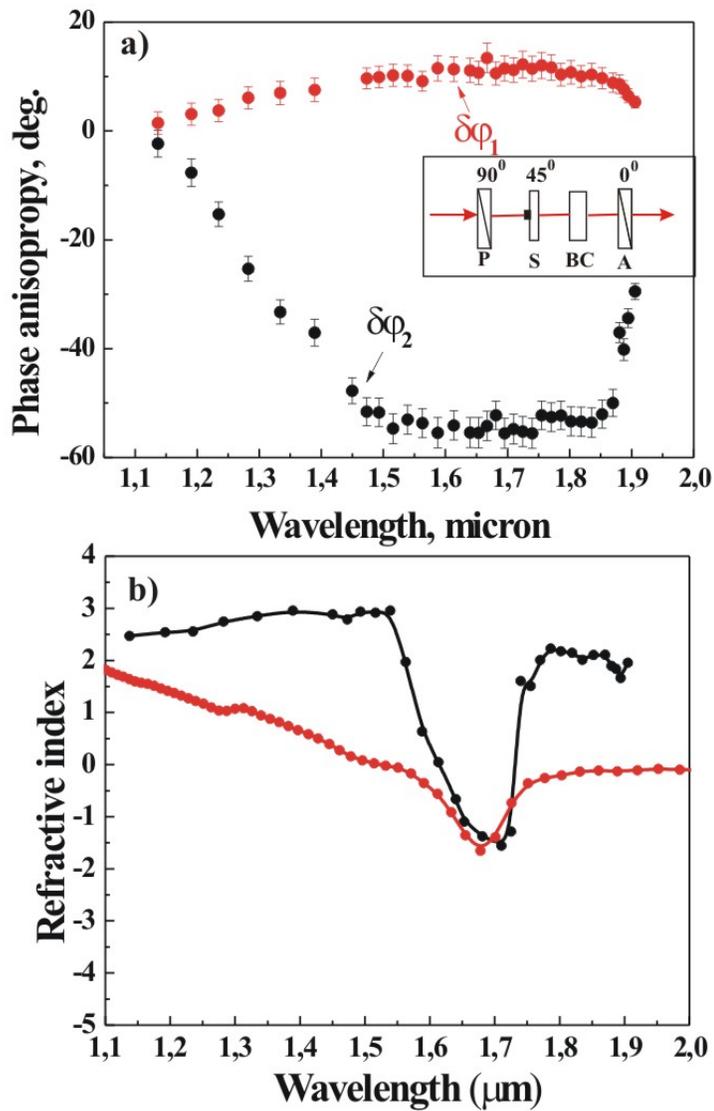

Fig.11 (a) Phase anisotropy for the transmitted (black curve) and reflected (red curve) probing light. The insert shows the scheme of the phase measurements for the transmitted light: P is polarizer, S is sample, BC is Babinet Compensator, A is analyzer. (b) The real part of the refractive index retrieved from the experimental data (black curve) and from the FDTD simulations (red curve).

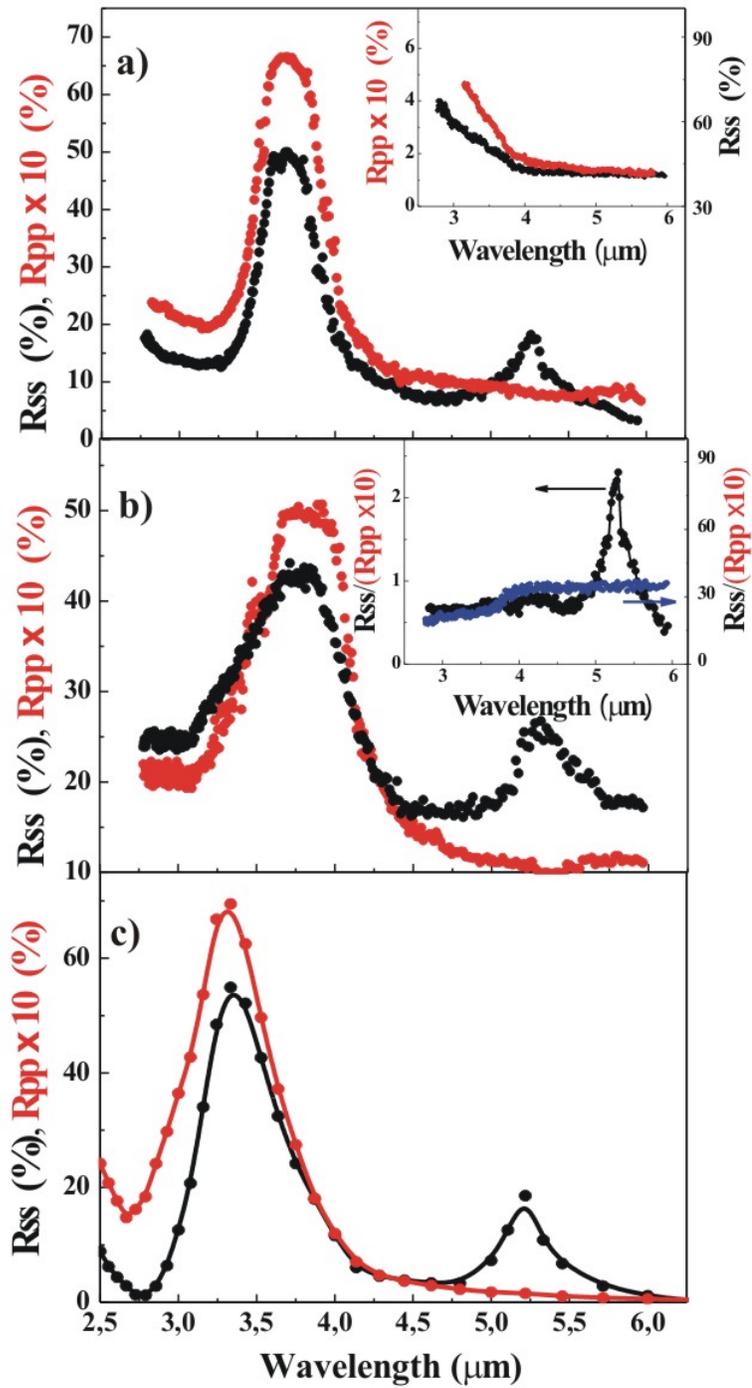

Fig. 12 (a) Reflection spectra of the LSR array on $Si_3N_4 + SiO_2$ substrate for s-in s-out (black curve) and p-in p-out (red curve) polarization configurations. The insert shows the reflection spectra for a clean $Si_3N_4+SiO_2$ substrate for s-in s-out (black curve) and p-in p-out (red curve) polarization configuration. (b) Reflection spectra of LSR array on Si + $SiO_2$ + $Si_3N_4$ substrate for s-in s-out (black curve) and p-in p-out (red curve) polarization configuration. The insert shows reflection spectra with s-in s-out polarizations normalized with respect to the p-in p-out polarizations from nanoimprinted LSRs on a $Si_3N_4$+$SiO_2$ substrate (black curve) and from the clean $Si_3N_4$ + $SiO_2$ substrate (blue curve). (c) Simulated reflection spectra for LSRs on a $Si_3N_4$+ $SiO_2$ substrate for s-in s-out (black curve) and p-in p-out (red curve) polarization configuration.